\documentclass[aps,prc,twocolumn,superscriptaddress,showpacs,floatfix]{revtex4}
\usepackage{graphicx}
\usepackage{epstopdf}

\begin{document}


\title{Quasiparticle structure and $\alpha$-decay scheme of nuclei along alpha-decay chain of $^{288}$Mc}

\author{L.A.Malov}
\affiliation{Joint Institute for Nuclear Research, 141980 Dubna, Moscow Region, Russia}
\author{N.Yu.Shirikova}
\affiliation{Joint Institute for Nuclear Research, 141980 Dubna, Moscow Region, Russia}
\author{A.N.~Bezbakh}
\affiliation{Joint Institute for Nuclear Research, 141980 Dubna, Moscow Region, Russia}
\author{E.A.~Kolganova}
\affiliation{Joint Institute for Nuclear Research, 141980 Dubna, Moscow Region, Russia}
\affiliation{Dubna State University, 141982 Dubna, Moscow Region, Russia}
\author{R.V.~Jolos}
\email[]{Email address: jolos@theor.jinr.ru}
\affiliation{Joint Institute for Nuclear Research, 141980 Dubna, Moscow Region, Russia}
\affiliation{Dubna State University, 141982 Dubna, Moscow Region, Russia}
\affiliation{Lomonosov Moscow State University, Dubna Branch, 141980 Dubna, Moscow Region, Russia}

\date{\today}

\pacs{21.10.Tg,23.20.Lv,23.20.-g,27.70.+q}

\begin{abstract} Recent experiments on $\alpha$-decay of odd-odd superheavy nuclei give an important
information on the structure of the low-lying states of these nuclei. For this reason it is interesting
to calculate the excitation spectra of these superheavy nuclei and compare the results with the experimental data.
The aim of this work is to calculate the excitation energies of the two-quasiparticle states of nuclei belonging to the $\alpha$-decay chain of $^{288}$Mc.
The approximation of the noninteracting quasiparticles based on the Woods-Saxon single
particle potentials is used.  Different sets of deformation parameters are considered.
The spectra of the low-lying two-quasiparticle  states are calculated.  The $\alpha$-decay
spectra of nuclei belonging to the $\alpha$-decay chain of $^{288}$Mc are obtained and compared
with the experimental data. A possibility of the $E1$ transitions in $^{276}$Mt and $^{272}$Bh following $\alpha$-decay of $^{288}$Mc is considered.
It is shown that the E1 transitions in $^{276}$Mt  can be related 
to the transition $\pi[505]9/2\rightarrow\pi[615]11/2$. In $^{272}$Bh the $E1$ transition can be related to the neutron single quasiparticle states.
\end{abstract}

\maketitle

\section{Introduction}
\label{first}
\indent

Experiments in which superheavy elements were synthesized in the complete fusion reactions with
$^{48}$Ca beam and actinide targets were successfully carried out in FLNR (Dubna), GSI (Darmstadt),
and LBNL (Berkeley) \cite{YuTs1,YuTs2,YuTs3,Hofmann,Loveland,Gregorich,Stavsetra,Rudolph1,YuTs4} producing superheavy nuclei with $Z=112-118$. This significantly expanded the region of nuclei whose properties can be investigated experimentally.
The investigations of transfermium elements have increased our
knowledge of single particle structure  and decay modes of heaviest nuclei
\cite{Parkhomenko1,Parkhomenko1a,Parkhomenko2,Adamian1,Herzberg,Hessberger,Hessberger1,Hessberger2,Shi1,Kortelainen,Shi2,Rudolph2,Bezbakh1}.
In recent years a set of experimental data on the structure of heaviest nuclei 
has been considerably increased because the 
experimental setups began to combine $\alpha$, $e^-$, and $\gamma$ spectroscopy \cite{Herzberg,Hessberger,Hessberger1,Hessberger2}.
In \cite{Rudolph1} the $\alpha$-decay chain of $^{288}$Mc were obtained. The structure of some low-lying states
of the odd-odd superheavy nuclei below $^{288}$Mc was clarified in $\alpha-\gamma$ coincidences. The $E1$-transitions in $^{276}$Mt were observed and the scheme of $E1$ transitions in $^{276}$Mt was suggested.
However, the scanty of available experimental information intensifies theoretical studies aimed at identifying various possibilities for their interpretation.
In this connection in Refs. \cite{Shi1,Kortelainen,Shi2} the shell structure of the odd-even nuclei in the $\alpha$-decay chain of $^{287}$Mc was studied and in \cite{Bezbakh1} the low-energy excitation spectra and $\alpha$-transition energies of odd-odd nuclei in the $\alpha$-decay chain of $^{288}$Mc was explored.

The products of the $\alpha$-decay chain of $^{288}$Mc combine nuclei from very quadrupole soft to deformed one.
For this reason the important meaning have the values of the equilibrium deformations and the amplitudes
of the zero point oscillations of the deformation parameter.

The aim of the present work is to calculate the excitation spectra of nuclei belonging to the $\alpha$-decay chain
of $^{288}$Mc and to identify possible $\alpha$- and $\gamma$-transitions. Especially, to indicate a possibility of $E1$ transitions between the low-lying states of $^{276}$Mt and $^{272}$Bh.

\section{Model}
\label{second}
\indent

In the present work the equilibrium deformations of nuclei belonging to the $\alpha$-decay chain of $^{288}$Mc are found with the
Nilsson-Strutinsky procedure based on the Woods-Saxon single particle potential. The potential parameters were adjusted to ensure that the calculated energies of the lowest one- and two-quasiparticle states closely matched those obtained using the two-center shell model (TCSM) potential \cite{Adamian2,Adamian3,Adamian4,Adamian5,Bezbakh1}. The following results have been obtained for the values of the quadrupole deformation: $^{288}$Mc -- $\beta_2=0.06$, $^{284}$Nh -- $\beta_2=0.02$, $^{280}$Rg -- $\beta_2=0.16$,
$^{276}$Mt -- $\beta_2=0.18$, $^{272}$Bh -- $\beta_2=0.26$, although hexadecapole deformation was also included in the calculations. These values of the quadrupole deformation differ from those given in \cite{Bezbakh1},
especially for $^{288}$Mc and $^{284}$Nh. The values of $\beta_2$ obtained in \cite{Bezbakh1} for other isotopes in the $\alpha$-decay chain of $^{288}$Mc are also slightly larger than those obtained in the present work. 
These differences are not surprising, as the calculations used different single-particle potentials, but
this indicates that the equilibrium deformation results are highly sensitive  to the choice of the single-particle potential for nuclei in this region of the chart of nuclides.

In the present approach the pairing is treated in the BCS approximation. Pairing interaction constants are determined by the experimental data on the odd-even mass differences for every odd-odd nucleus separately.

Consider the spectra of the single quasiparticle states obtained with deformations listed above. The most interesting is the spectrum of $^{276}$Mt where the $E1$ transitions between the low-lying states have been observed experimentally~\cite{Rudolph1}. In this part of the nuclide chart there is a very limited number of pairs of proton and neutron single particle levels connected by the allowed $E1$ transitions.

The spectrum of the low-lying states of $^{276}$Mt with odd proton in [615]11/2, [521]1/2, and [505]9/2 single quasiparticle states is shown in Fig.1. It is seen that there are several pairs of two-quasiparticle states related by  the $E1$ transition. The excited states which can be populated by the $E1$ transitions have excitation energies not exceeding 230 keV. These $E1$ transitions connect [505]9/2 and [615]11/2 proton single quasiparticle states.

\begin{figure}
\includegraphics{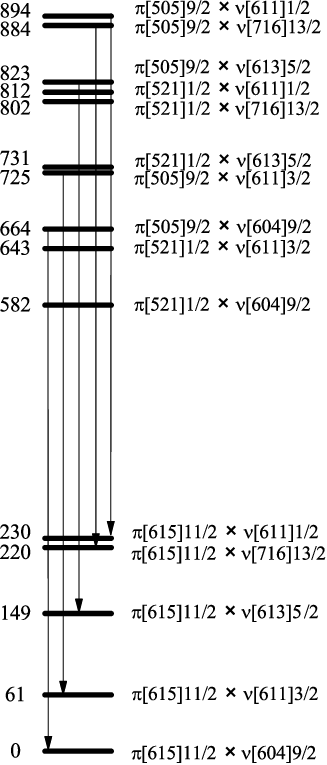}
\vspace*{-0.0cm}
\caption{\label{Fig1} The low-lying two-quasipartical states of $^{276}$Mt calculated with deformation $\beta_{20}=0.18$ and $\beta_{40}=-0.064$. Only states with odd proton in [615]11/2, [521]1/2, and [505]9/2 single quasiparticle states are indicated.
}
\end{figure}

The following question that needs to be investigated is the possibility of population in $\alpha$-decay of $^{288}$Mc those states of $^{276}$Mt which can de-excited by $E1$ transitions to low-lying states, i.e. states with odd proton in [505]9/2 single quasiparticle state. The analysis of the calculated $\alpha$-decay spectra of nuclei between $^{288}$Mc and $^{276}$Mt have shown that the states of $^{276}$Mt which can decay by $E1$ transition to the low-lying states  are not populated.
The reason is a large difference in the single particle spectra of nuclei included in the $\alpha$-decay chain of $^{288}$Mc, associated with the difference in their equilibrium deformations. The two-quasiparticle states of $^{288}$Mc with odd proton occupying [505]9/2
single quasiparticle state are located above 2150 keV. In $^{284}$Nh these states are located above 2000 keV. Such high excitation energies excludes a population with sufficient probability  in $\alpha$-decay of $^{288}$Mc and $^{284}$Nh of two-quasiparticle states with odd proton in [505]9/2
single quasiparticle state. As a consequence, the states of $^{276}$Mt from which $E1$ transitions to the low-lying states are possible can not be populated.

The above results are a consequence of the large difference in quadrupole deformations of $^{288}$Mc and $^{284}$Nh on the one side and $^{280}$Rg, $^{276}$Mt, and $^{272}$Bh on the other side. For this reason we decided to take as the basis the variant  of the quadrupole deformation parameter with smaller variation with the mass number $A$. We have chosen the results published in \cite{Jachimowicz}.
However, we have changed  the parameters given in \cite{Jachimowicz} for $^{284}$Nh so as to have a more smooth change in the deformation parameters along the $\alpha$-decay chain of $^{288}$Mc. Thus, we fixed for $^{284}$Nh: $\beta_{20}=0.09$ and $\beta_{40}=-0.05$.

Consider, at first, the excitation spectrum of $^{276}$Mt in which there are several two-quasiparticle states related by $E1$ transition (see Fig.2). As above, these $E1$ transitions connect [505]9/2 and [615]11/2 odd proton single quasiparticle states. In this variant, the two-quasiparticle states of $^{276}$Mt   decaying by $E1$ transitions to the ground and low-lying states have excitation energies around 1 MeV. 
The excited states with odd proton in [615]11/2 single quasiparticle state which are populated by $E1$ transition have excitation energies below 194 keV. 
Correspondingly, these states will populate in $\alpha$-decay the states of $^{272}$Bh with odd proton in [615]11/2 single quasiparticle state.
Because excited two-quasiparticle states of $^{272}$Bh with odd proton in [505]9/2 single quasiparticle states are located much higher (above 2000 keV) it excludes a possibility of the $E1$ transition
[615]11/2$\rightarrow$[505]9/2 of odd proton in this nucleus. There is, however, a possibility of the $E1$ transition between the states with odd proton in [624]9/2 and [514]7/2 single quasiparticle states. In this case,  difference in the excitation energies of the states connected by $E1$ transition is around 500 keV. 
\begin{figure}
\includegraphics{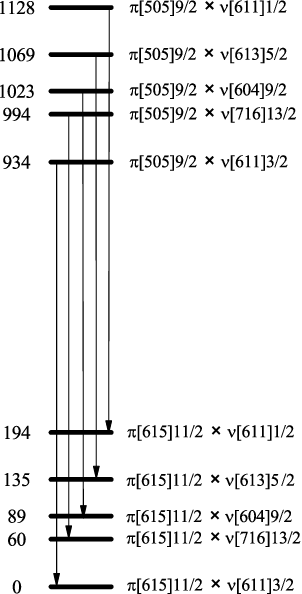}
\vspace*{-0.0cm}
\caption{\label{Fig2} The low-lying two-quasipartical states of $^{276}$Mt calculated with deformation $\beta_{20}=0.21$ and $\beta_{40}=-0.09$ and having odd proton in one of the following single quasiparticle states:  [615]11/2 and [505]9/2.
}
\end{figure}

The following question that needs to be investigated is a possibility of population in $\alpha$-decay of $^{288}$Mc those states of 
$^{276}$Mt from which $E1$ transitions to the low-lying states come.
The excited states which should be populated in $^{276}$Mt to create $E1$ transitions in this nucleus are located at 934 keV, 994 keV, 1023 keV, 1069 keV, and 1128 keV. They are $\pi$[505]9/2$\times\nu$[611]3/2, $\pi$[505]9/2$\times\nu$[716]13/2, $\pi$[505]9/2$\times\nu$[604]9/2, $\pi$[505]9/2$\times\nu$[613]5/2, and $\pi$[505]9/2$\times\nu$[611]1/2. Taking into consideration only allowed cases of
$\alpha$-decay, i.e. cases with unchanged  quantum numbers of odd proton and odd neutron we consider in nuclei belonging to the $\alpha$-decay chain of 
$^{288}$Mc only positions of the two-quasiparticle states mentioned above. From the calculated excitation spectrum of $^{288}$Mc we see that these states are located in $^{288}$Mc at sufficiently high excitation energies: from 2146 keV to 3198 keV. In addition, these states decay by the strong $E1$ transitions ($\tau_{1/2}=1.4\cdot 10^{-12}$ sec) to the lower lying states with odd proton occupying 
[615]11/2 single quasiparticle states. It follows from these facts that the states of $^{276}$Mt with odd proton occupying [505]9/2 single quasiparticle states are not populated in the case of the considered variant of the deformation parameters. The $\alpha$-decay scheme of $^{288}$Mc containing states discussed above is shown in Fig.3.

\begin{figure*}
\includegraphics{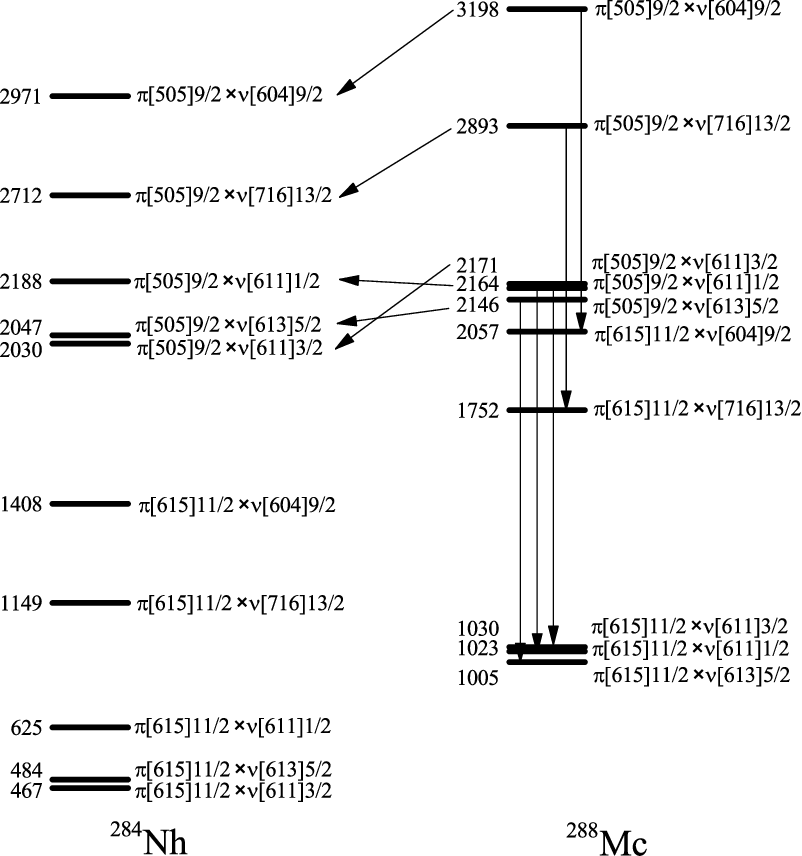}
\vspace*{-0.0cm}
\caption{\label{Fig3} $\alpha$-decay scheme of $^{288}$Mc with deformation parameters: $^{288}$Mc ($\beta_{20}$=0.08) and $^{284}$Nh ($\beta_{20}$=0.09).
}
\end{figure*}

Our analysis of the spectra of the excited states of nuclei included in the $\alpha$-decay chain of $^{288}$Mc obtained when considering various variants of deformations of the average nuclear potential have shown that the best agreement with experimental data,  especially with data on $E1$ transitions in $^{276}$Mt, is achieved at the following values of deformation parameters: $^{288}$Mc ($\beta_{20}=0.14$, $\beta_{40}=-0.049$), $^{284}$Nh ($\beta_{20}=0.16$, $\beta_{40}=-0.049$), $^{280}$Rg ($\beta_{20}=0.16$, $\beta_{40}=-0.049$), $^{276}$Mt ($\beta_{20}=0.20$, $\beta_{40}=-0.063$), and $^{272}$Bh ($\beta_{20}=0.20$, $\beta_{40}=-0.063$). These deformation parameters  for $^{272}$Bh, $^{276}$Mt, and $^{280}$Rg nuclei are quite close to those given in \cite{Jachimowicz}. However, for 
$^{288}$Mc and $^{284}$Nh isotopes the values of the $\beta_{20}$ parameter are noticeably larger than  obtained in \cite{Jachimowicz}. Note that the values of $\beta_{20}$ for $^{288}$Mc and $^{284}$Nh obtained in the previous variants characterize these nuclei as spherical or transitional. We suppose that the values of $\beta_{20}=0.14$ and $\beta_{20}=0.16$ given in the last of the considered versions, rather characterize the amplitudes of zero point oscillations of $\beta_{20}$ in these nuclei, than the equilibrium values of this quantity. For example, the experimental value of $\beta_{20}$ for $^{222}$Rn extracted from the value of $B(E2;0^+_1\rightarrow 2^+_1)$
is 0.14, while $R_{4/2}=2.4$ and $R_{6/2}=4.1$ ratios are characteristic of nuclei close to spherical.
The experimental value of $\beta_{20}$ for $^{222}$Ra 
is 0.19, while $R_{4/2}=2.71$ and $R_{6/2}=4.95$ ratios are typical for transitional nuclei. The results of calculation of the equilibrium 
$\beta_{20}$ values obtained in \cite{Bezbakh1} also indicate that $^{288}$Mc and $^{284}$Nh should be attributed to spherical or transitional nuclei.

Therefore, the optimal model for describing the properties of these nuclei should include collective quadrupole oscillations via Bohr Hamiltonian with potential that permits a large amplitude of shape oscillations. At the same time,  single particle states should be presented by the spherical shell model Hamiltonian, but the full Hamiltonian includes a term describing  a coupling of  single particle motion with the quadrupole oscillations of nuclear shape. However, experimental information on even-even and odd nuclei from this region of the nuclide chart is so scarce that an attempt to construct the Hamiltonian described above would encounter a significant uncertainties.

In view of the above, we consider, for simplicity, $^{288}$Mc and $^{284}$Nh 
in the framework of the scheme of other nuclei -  $^{280}$Rg, $^{276}$Mt, and $^{272}$Bh, however, with quadrupole deformation parameters which, in fact, are mean square shape fluctuation amplitudes. It is acceptable, since shape fluctuation amplitudes characterize region in the deformation parameter space where the nucleus wave function is concentrated. This suggestion is consistent with an absence of abrupt changes in $\alpha$-decay characteristics of the nuclei in question. In addition, $E1$ transitions between low-lying states in $^{276}$Mt and $^{272}$Bh observed experimentally indicate a similarity of the schemes of the single particle levels in these nuclei, since in this part of the nuclide chart there is a very limited number of pairs of proton and neutron single particle levels connected by the allowed $E1$ transitions. This is confirmed also by the results of calculations of the low-lying level schemes  in the odd proton nuclei adjacent to the considered ones, performed in different approaches
\cite{Parkhomenko1,Shi1,Rudolph2,Kuzmina}.

The main task of this work  is the calculation of the spectra of the excited states of nuclei belonging to the $\alpha$-decay chain of $^{288}$Mc and the identification of possible $\alpha$- and competing $\gamma$-transitions. It is clear that taking into account the assumptions made above, the results obtained can only be indicative in nature and can only be considered  as a reference point.

Consider now the consequence of the last variant of the set of the equilibrium deformations  suggested as the optimal for description of the experimental data.  Consider at first, the excitation spectrum of $^{276}$Mt (Fig.4).

\begin{figure}
\includegraphics{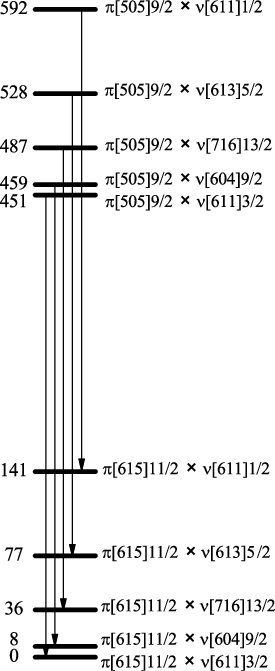}
\vspace*{-0.1cm}
\caption{\label{Fig4} The low-lying two-quasiparticle states of $^{276}$Mt calculate with deformation parameters $\beta_{20}=0.20$ and
$\beta_{40}=-0.063$. Only states with odd proton occupying single quasiparticle states [615]11/2 and [505]9/2 are indicated.}
\end{figure}
It is seen that there are five pairs of two-quasiparticle states related by $E1$ transitions. The excited states which are populated by $E1$ transition have excitation energies below 150 keV.  As in the case of the other sets of deformation parameters these $E1$ transitions connect
[505]9/2 and [615]11/2 proton single quasiparticle states. Mention, that in $^{276}$Mt neutron single quasiparticle states
[611]3/2, [604]9/2, [716]13/2 and [613]5/2 are very close to each other in excitation energy. They are concentrated in the energy interval 77 keV. The single quasiparticle neutron state [611]1/2 is also close to them. For this reason, we do not consider electromagnetic transitions between these states. 

Let us investigate for this variant of deformation parameters a possibility of population in $\alpha$-decay of $^{288}$Mc those states of 
$^{276}$Mt from which $E1$ transitions to the low-lying states come. The excited states of $^{276}$Mt which should be populated to create $E1$ transitions are located at 451 keV, 459 keV, 487 keV, 528 keV, and 592 keV. They are $\pi$[505]9/2$\times\nu$ [611]3/2, 
$\pi$[505]9/2$\times\nu$ [604]9/2, $\pi$[505]9/2$\times\nu$ [716]13/2, $\pi$[505]9/2$\times\nu$ [613]5/2, and $\pi$[505]9/2$\times\nu$ [611]1/2, correspondingly. Taking into account only cases of $\alpha$-decay with unchanged quantum numbers of odd proton and odd neutron we obtain $\alpha$-decay scheme of the considered nuclei presented in Fig.5.

\begin{figure*}[htpb]
\includegraphics{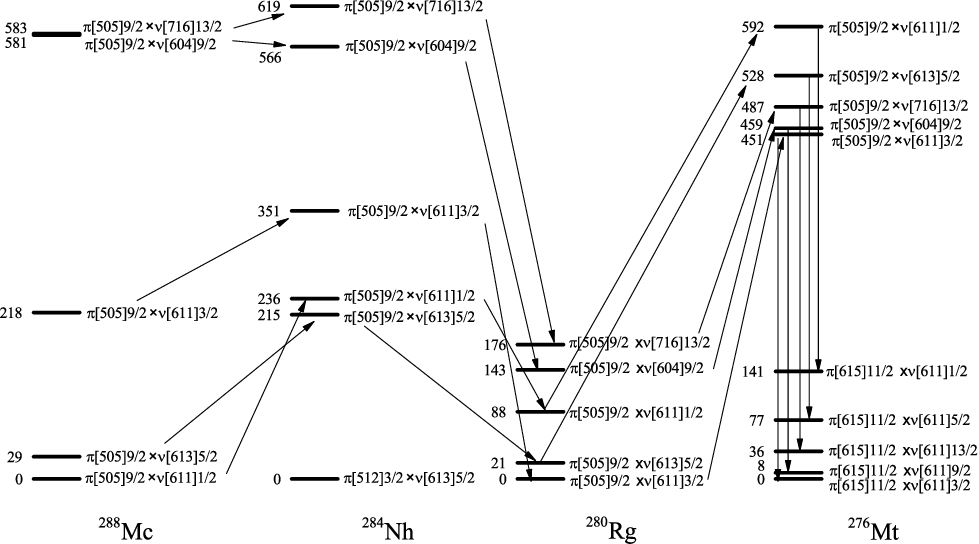}
\vspace*{-0.0cm}
\caption{\label{Fig5} $\alpha$-decay scheme of $^{288}$Mc. Only states with odd proton occupying [505]9/2 single quasiparticle state are shown. Excitation energies are given in keV.
}
\end{figure*}

From the calculated excitation spectrum of $^{288}$Mc we see that the excited states with quantum numbers which should be populated in $^{276}$Mt are located in $^{288}$Mc at excitation energies from the ground state up to 583 keV. Consider a possibility of deexcitation of some of these states to the low-lying states.
The neutron single quasiparticle states [613]5/2 and [611]1/2 are very close in energy. So, we can neglect a possibility of the E2 transition between $\pi$[505]9/2$\times\nu$ [613]5/2 and $\pi$[505]9/2$\times\nu$ [611]1/2.
 However, E2 transition between $\pi$[505]9/2$\times\nu$ [604]9/2 and $\pi$[505]9/2$\times\nu$ [613]5/2 is more fast than $\alpha$-decay of $\pi$[505]9/2$\times\nu$ [604]9/2 state. Therefore, population of $\pi$[505]9/2$\times\nu$ [604]9/2 in
$^{276}$Mt is under the question. Neutron single quasiparticle states [611]3/2 and [613]5/2 form pseudospin doublet. For this reason, we can neglect E2 transition  between these states.

In $^{284}$Nh the proton single quasiparticle states [615]11/2 and [505]9/2 are very close in energy (11 keV difference). So, we can neglect $E1$ transitions between these states. In $^{280}$Rg proton single quasiparticle states [615]11/2 and [505]9/2 are also very close in energy (2 keV difference) and in addition [505]9/2 is lower in energy than [615]11/2. Thus, the excited states of $^{276}$Mt which decay by $E1$ transition to the low-lying states can be populated in $\alpha$-decay of $^{288}$Mc  in the considered variant of deformation parameters.

In $\alpha$-decay of $^{276}$Mt two-quasiparticle states of $^{272}$Bh with odd proton in the state [615]11/2 are populated. They lies lower in energy than two-quasiparticle states with odd proton in [505]9/2 state. Therefore,  $E1$ transitions in $^{272}$Bh associated with a transition between  the odd proton state are excluded. In principle, in $^{272}$Bh  $E1$ transition associated with a change in the state of the odd neutron [725]11/2$\rightarrow$[615]9/2 is possible. For example, $\pi$[615]11/2$\times\nu$[725]11/2$\rightarrow\pi$[615]11/2$\times\nu$[615]9/2. However, the state $\pi$[615]11/2$\times\nu$[725]11/2 is located in $^{276}$Mt at the excitation energy 1016 keV. This makes it unlikely that the corresponding state will be populated in $^{272}$Bh. Thus, in the case of this set of deformation parameters it is not possible to explain an appearance of $E1$ transition in $^{272}$Bh.

\section{Gallaher-Moszkowski  splitting}

One of the most significant effects associated with the residual neutron-proton interaction in odd-odd deformed nuclei is the Gallaher-Moszkowski (GM) splitting \cite{GM}. This effect has been studied in several papers~\cite{Pyatov,Jones,Alexa,Jain,Nosec}, where both zero-range and finite-range nuclear forces were considered. The energy splitting calculated for both types of forces agrees with corresponding excitation energies, which confirms the validity of the GM model.

To estimate the GM-splitting in
the odd-odd nuclei belonging to the $\alpha$-decay chain of $^{288}$Mc we have used a simple variant of calculations presented by N.I.Pyatov \cite{Pyatov}.
 Following this work  we obtain the following results.  If with positive of the angular momentum projections on the symmetry axis of proton ($\Omega_{\pi}$) and neutron ($\Omega_{\nu}$) the spins of the nucleons are parallel, then the lowest
will be the state with $K=\Omega_{\pi}+\Omega_{\nu}$. The energy splitting $\Delta E_{12}\equiv E_{12}(K=\Omega_{\pi}-\Omega_{\nu})-E_{12}(K=\Omega_{\pi}+\Omega_{\nu})$
will be equal to $-2\alpha A$, where
\begin{eqnarray}
\label{GM1}
A=F^{(0)}\sum_{l_{\pi},l_{\nu}}(2l_{\pi}+1)(2l_{\nu}+1)d^2_{l_{\pi}\Lambda_{\pi}}d^2_{l_{\nu}\Lambda_{\nu}}\nonumber\\
\times\sum_L\frac{1}{(2L+1)}\left(C^{L0}_{l_{\pi}0   l_{\nu}0}C^{L\Lambda_{\pi}-\Lambda_{\nu}}_{l_{\pi}\Lambda_{\pi}   l_{\nu}-\Lambda_{\nu}}\right)^2.
\end{eqnarray}
Above $d_{l\Lambda}$ are expansion coefficients of the Nilsson states given in the spherical basis, $F^{(0)}$ is a well known radial integral which appears in presentation of the Surface Delta Interaction \cite{Moszkowski}. We have taken $F^{(0)}$ to be equal to $-5.7 A^{-2/3}$MeV \cite{Heussler}.

If with positive $\Omega_{\pi}$ and $\Omega_{\nu}$ the spins of the nucleons are antiparallel, then the lowest
will be the state with $K=|\Omega_{\pi}-\Omega_{\nu}|$ and $\Delta E_{12}=-2\alpha B$, where
\begin{eqnarray}
\label{GM4}
B=F^{(0)}\sum_{l_{\pi},l_{\nu}}(2l_{\pi}+1)(2l_{\nu}+1)d^2_{l_{\pi}\Lambda_{\pi}}d^2_{l_{\nu}\Lambda_{\nu}}\nonumber\\
\times\sum_L\frac{1}{(2L+1)}\left(C^{L0}_{l_{\pi}0   l_{\nu}0}C^{L\Lambda_{\pi}+\Lambda_{\nu}}_{l_{\pi}\Lambda_{\pi}   l_{\nu}+\Lambda_{\nu}}\right)^2.
\end{eqnarray}
These expressions are obtained using a simple proton-neutron interaction
\begin{eqnarray}
\label{GM2}
V_{\pi\nu}=-4\pi g\delta({\vec r}_1-{\vec r}_2)(1-\alpha+\alpha {\vec \sigma}_{\pi}\cdot{\vec \sigma}_{\nu}).
\end{eqnarray}
In the calculations below $\alpha$ is taken to be equal to 0.5 \cite{Pyatov}.

We have calculated $\Delta E_{12}$ for several proton-neutron pairs. The results are the following:\\
$^{288}$Mc: $\pi[512]5/2\times\nu[611]3/2$ (ground state configuration). The lowest member of the doublet has $K=0$ and $\Delta E_{12}=66$ keV.\\
$^{280}$Rg: $\pi[505]9/2\times\nu[716]13/2$ (96 keV). The lowest member of the doublet has $K=1$ and $\Delta E_{12}=96$ keV.\\
$^{276}$Mt: $\pi[615]11/2\times\nu[604]9/2$ (ground state configuration). The lowest member of the doublet has $K=10$ and $\Delta E_{12}=115$ keV.\\
$^{272}$Bh: $\pi[521]1/2\times\nu[611]3/2$ (ground state configuration). The lowest member of the doublet has $K=1$ and $\Delta E_{12}=28$ keV.

\section{Summary}

The excitation spectra and decay schemes of nuclei in the $\alpha$-decay chain of $^{288}$Mc have been investigated using three sets of deformation parameters. The first one has been obtained in the present work basing on the Woods-Saxon single particle potential  and the Nilsson-Strutinsky procedure. The second set was taken from \cite{Jachimowicz}, however, we have changed $\beta_{20}$ for $^{284}$Nh  to get a more smooth change of $\beta_{20}$ along the $\alpha$-decay chain of $^{288}$Mc. The third set of deformation parameters includes the  values of $\beta_{20}$ 
for $^{280}$Rg, $^{276}$Mt, and $^{272}$Bh close to those taken in the second variant. However, the values of $\beta_{20}$ have been significantly increased for $^{288}$Mc and $^{284}$Nh. The analysis of the spectra of the excited states of nuclei included in the $\alpha$-decay chain of $^{288}$Mc have shown that the improvement of the agreement of the results of calculation with the experimental data requires just this modification of the values of $\beta_{20}$. This change was justified by the fact that 
small values of $\beta_{20}$ obtained in calculations for
$^{288}$Mc and $^{284}$Nh being atypical for deformed nuclei indicate that these nuclei are rather spherical or transitional. In this case, the amplitude of the mean square fluctuations of $\beta_{20}$ effectively plays the role of deformation parameters and we put them equal to the typical values of the mean square fluctuations of $\beta_{20}$ for heavy nuclei.

The results obtained show that only in the case of the third variant of deformation parameters  it is possible to explain observation of $E1$ transition in $^{276}$Mt. Strictly saying, it means that $^{288}$Mc and $^{284}$Nh should be considered  based on the Bohr collective Hamiltonian with particle - core coupling interaction.

In order to have an idea of the amount of splitting between proton-neutron states in deformed nuclei with parallel and antiparallel  couplings  we calculated the Gallaher-Moszkowcki splitting for a number of states in nuclei under consideration.

\end{document}